# EXPOSED: An occupant exposure model for confined spaces to retrofit crowd models during a pandemic

Enrico Ronchi[1], Ruggiero Lovreglio[2]

[1]Department of Fire Safety Engineering, Lund University, Lund, Sweden
[2]School of Built Environment, Massey University, Auckland, New Zealand

## Abstract

Crowd models can be used for the simulation of people movement in the built environment. Crowd model outputs have been used for evaluating safety and comfort of pedestrians, inform crowd management and perform forensic investigations. Microscopic crowd models allow the representation of each person and the obtainment of information concerning their location over time and interactions with the physical space/other people. Pandemics such as COVID-19 have posed several questions on safe building usage, given the risk of disease transmission among building occupants. Here we show how crowd modelling can be used to assess occupant exposure in confined spaces. The policies adopted concerning building usage and social distancing during a pandemic can vary greatly, and they are mostly based on the macroscopic analysis of the spread of disease rather than a safety assessment performed at a building level. The proposed model allows the investigation of occupant exposure in buildings based on the analysis of microscopic people movement. Risk assessment is performed by retrofitting crowd models with a universal model for exposure assessment which can account for different types of disease transmissions. This work allows policy makers to perform informed decisions concerning building usage during a pandemic.

## Highlights

- Crowd models are useful tool to perform risk assessments in buildings
- EXPOSED, an occupant exposure model is presented
- EXPOSED calculates occupant exposure based on crowd model outputs
- Policy makers can use EXPOSED to perform informed decisions on building use

## Nomenclature

$C_k$ : cumulative exposure to a given number of people $k$

$e_{tf}$ : the $k$ number of agents $j$ to which each individual agent $i$ is exposed at the time-step $t_f$

$e_{tq}$ : the $k$ number of agents $j$ to which each individual agent $i$ is exposed at the time-step $t_q$

$\boldsymbol{E^i}$ : set of number of agents $j$ each individual agent $i$ is exposed to at each time-step $t_q$

$\boldsymbol{E_t^i}$ : matrix of occupant exposure of each individual agent $i$ to other agents $j$ at each time-step $t_q$

$G$ : global assessment of exposure for the total time $t_f$ spent by all $n$ agents in the building

$i$ : individual agent under consideration

$j$ : individual agent to which $i$ is exposed to

$k$ : number of $j$ agents an individual agent $i$ is exposed to in a given time-step $t_q$

$m$ : generic maximum number of agents to which agents are exposed to

$n$ : maximum number of agents in the building

$R_i$ : social distance radius of an individual agent $i$

$\boldsymbol{t}$ : time-step

$t_f$ : final time-step at which all agents have left the building

$t_q$ : generic time-step in the simulation

$t_{k,max}^i$ : maximum time of exposure for each simulated agent $i$ to each number of $k$ agents

$t_q^i$ : the number of time-steps $t_q$ in which each agent $i$ is interacting with a discrete number of people $k$

$T_k^i$ : time of exposure of each agent $i$ to a given number of $k$ agents

$T_k$ : distribution of exposure times to a given discrete number of agents $k$

$\alpha_{ij}$ : polar coordinate defining the position of the agent $j$ in the polar space in relation to each agent $i$

$\beta_{ij}$ : orientation of the agent $j$ in the polar space defined by the agent $i$

$\gamma_k$ : factor which increases the exposure in relation to the value of $k$

$\mu_k$ : average value of the $T_k$ distribution of exposure times to a given discrete number of agents $k$

$\rho_{ij}$ : polar coordinate defining the position of the agent $j$ in the polar space in relation to each agent $i$

$\sigma_k^2$ : variance of the $T_k$ distribution of exposure times to a given discrete number of agents $k$

## 1. Introduction

Any stakeholder dealing with crowds is greatly affected by a pandemic. Events involving large crowds have been cancelled or postponed worldwide, access to public buildings has been restricted as different mitigation measures have been adopted around the world to decrease physical interactions among people during the COVID-19 pandemic (Anderson et al., 2020). The measures to face pandemics adopted by policy makers ranged from compulsory lockdowns to recommendations on social distancing (also called physical distancing). One key aspect sticks out in the current situation, i.e. the inconsistency in the adopted measures is evident. For instance, to the date this paper was written (May 2020), social distancing recommendations are diverse in different countries and rapidly change over time, e.g., 1 person/4 $m^2$ (Australia), 1 m (Philippines, Qatar) or 2 m outside home (Canada, UAE, UK, South Korea), or 1-2 m (New Zealand, Italy) (Italian Government, 2020; Korean Ministry of Economy and Finance, 2020; Movement Strategies, 2020). This raises questions on their implications on space usage (Honey-Roses et al., 2020), how policy makers have grounded their decisions and what type of models are currently available to support these decisions. Moreover, current data do not fully support the 1-2 m rule for spatial separation assumed for travelling of large droplets (Bahl et al., 2020) which has been issued in the airborne precautions for the COVID-19 pandemic in the guidelines by the World Health Organization (WHO, 2020). This issue highlights the need for flexible models able to use different assumptions on disease transmission mechanisms.

Current research supporting decision makers is mostly based on macroscopic epidemiological models, among which the most used are different variations of the SIR model. This model is an epidemiological tool that estimates the number of people infected in given conditions over time. The SIR model divides the population into three types, namely (1) susceptible, S; (2) infectious, I; and (3) recovered/removed, R. The SIR model is based on earlier analytical approaches developed to study disease spread back in the 1920ies (Kermack & McKendrick, 1927). The current field of mathematical epidemiology adopts a set of differential equations which generally considers also a (4) exposed class, E, to create the so-called SEIR model (Anderson et al., 1992). Several other macroscopic epidemiological models have been developed and used, including stochastic transmission models (Kucharski et al., 2020) and mean-field epidemiological models (Giordano et al., 2020).

The use of macroscopic epidemiological models provides the great benefit to be applicable at large scales and give vital information to decision makers as they yield predictions of the spread of disease. Nevertheless, their great limitation is that they may not fully consider the mobility patterns of individuals and the heterogeneity in their interactions. This issue has been raised in the largest conference in the pedestrian and evacuation dynamics community back in 2012 (Johansson & Goscè, 2014) and has initiated discussions on coupling the field of crowd dynamics and modelling with epidemiological modelling (Goscè et al., 2014). The impact of heterogeneity in population and mobility patterns have been so far addressed mostly at a macroscopic scale, especially in transportation environments (Goscé & Johansson, 2018; Meloni et al., 2011; Saberi et al., 2020) in which transport networks are investigated to consider them while modelling the spread of disease. Nevertheless, current disease-spreading models do not explicitly represent the space usage of pedestrians at a microscopic scale, thus making difficult to assess at a building level what the impact of a social distance measure could be.

Microscopic crowd models, such as continuous models (Helbing et al., 2000; Thompson & Marchant, 1995) and discrete models (Lovreglio et al., 2015), have been used to represent the individual movement of pedestrians in confined spaces and provide – among other outputs – information concerning the trajectories of pedestrians in space over time (i.e. this is generally in the form of the parametric equations of each pedestrian). Crowd models have been used so far mostly to ensure comfort and safety of pedestrians and identify crowd management solutions to optimize movement flows and reduce waiting times (Bellomo & Gibelli, 2016; Johansson, 2008). It appears evident that the simulation of people movement at a microscopic scale could provide a great help to decision makers for risk assessment in case of disease spreading in a confined space. During the COVID-19 pandemic, some of the most known and used crowd models (Lovreglio et al., 2019) have released new features aimed at microscopic modelling of people movement considering social distancing or counting the interactions between pedestrians in a given social distance radius. While these features are useful to evaluate space usage, given the lack of knowledge concerning the current spread of disease, they do not allow a comprehensive quantitative understanding of the impact of different measures on building occupant exposure. It is therefore crucial to develop a general occupant exposure model which could be used to retrofit any type of microscopic crowd model and that is able to produce quantitative outputs for risk assessment. In turn, crowd models do not currently provide dedicated comprehensive sub-models which can be directly used to perform this type of analysis.

In this paper we propose a modelling solution to retrofit any type of microscopic crowd model for pandemic risk assessment studies. To achieve this goal, we introduce a universal occupant exposure model which could be used to retrofit any existing microscopic crowd model able to track the trajectories of pedestrians over time in a confined space. The ground for the development of the model has been the review of the current measures adopted to reduce risk transmission worldwide (e.g. social distancing measures), possible disease transmission mechanisms in confined spaces and the analysis of the characteristics of microscopic crowd modelling tools.

## 2. Modelling assumptions on occupant exposure

The proposed model is developed using a general formulation rather than relying on a specific type of disease transmission. This has been done so that it could be adapted for different types of disease transmissions, i.e. based on different distances between pedestrians, types of contact (e.g. airborne vs droplet, angle of contact, etc.), reference points (e.g. pedestrian face, nose, shoulder, etc.) and time of exposure. This is deemed appropriate for new pandemics such as COVID-19 in which there is no conclusive understanding on the disease transmission (Lewis, 2020), thus providing the flexibility to update the model results in case new insights on the pandemics are available.

Since the severe acute respiratory syndrome coronavirus 2 (SARS-CoV-2) transmission is currently not fully understood, one of the current assumptions is that it can be considered similar to the severe acute respiratory syndrome coronavirus (SARS-CoV-1) in which the spread was found to be by (1) physical contact, (2) droplets, and (3) airborne routes (Yu et al., 2004). Given the uncertainty in disease transmission mechanisms during the COVID-19 pandemic, five different assumptions on occupant exposures are here suggested as possible mechanisms to adopt in the proposed model. These are presented graphically in Figure 1 and described in Sections 2.1-2.5.

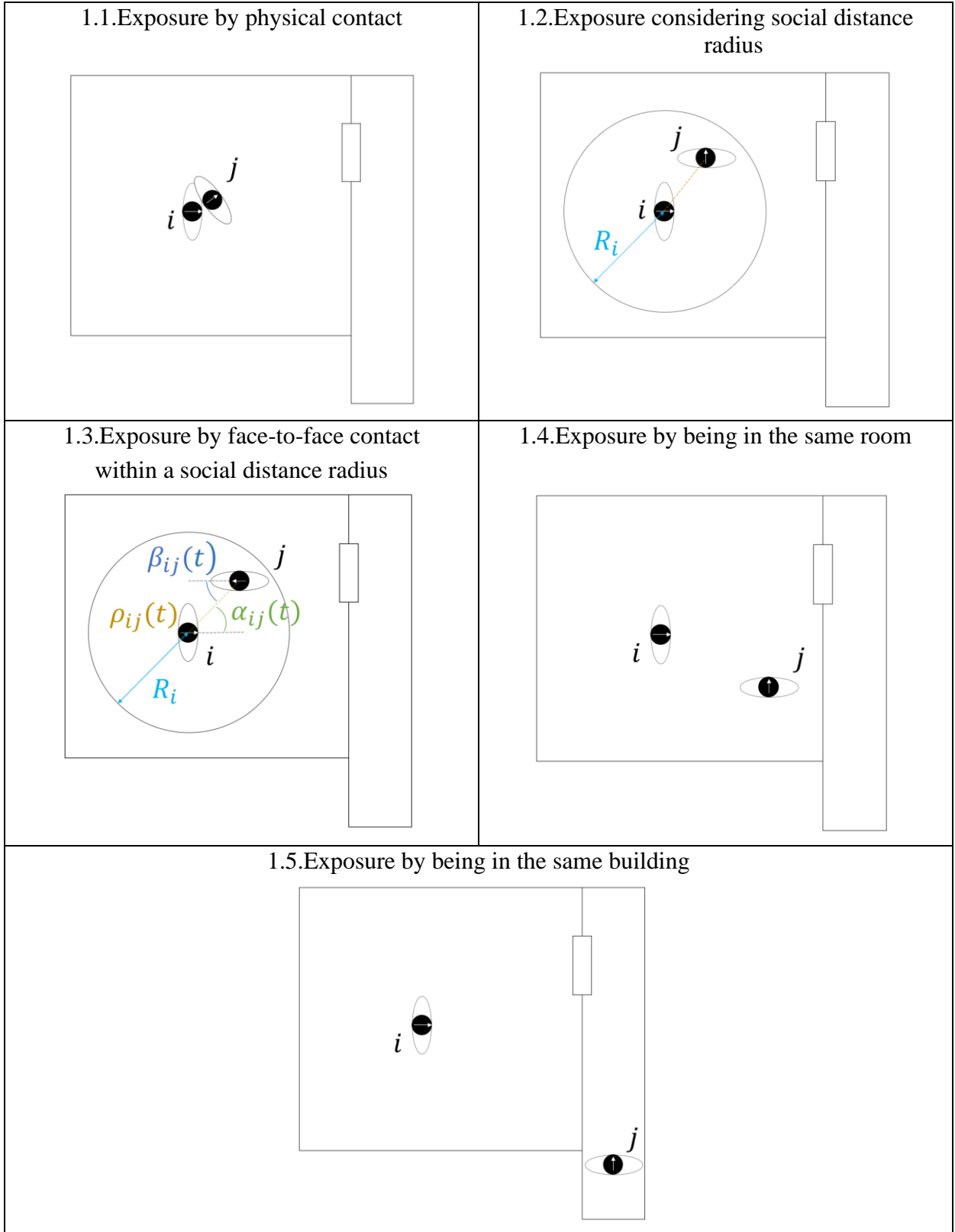

*Figure 1. Examples of assumptions on occupant exposure considering a room and a corridor connected by a door and two simulated pedestrians within them.*

### 2.1. Exposure by physical contact

Occupant exposure in a confined space is here accounted for when people are in direct physical contact to each other (i.e. pedestrians "touch each other"). It should be noted that most current crowd models generally assume rigid bodies (Duives et al., 2013), i.e. non-deformable pedestrian bodies which allow the detection of collision and contacts between agents. Therefore, this is practically implemented considering the representation of the agents (often represented in crowd models as circles or ellipses) and exploring that their coordinate in space overlap within a certain radius corresponding to their dimensions. The example in Figure 1.1 shows one agent $i$ and one agent $j$ that are in physical contact within a room. This assumption requires from a crowd model the information concerning the pedestrian trajectories over time and the dimension of the agents.

### 2.2. Exposure considering social distance radius

Occupant exposure in a confined space assumes here the number of people in a social distance radius defined by the user (e.g., 1 or 2 m). The centre of the social distance radius can be assumed to be the centre of the modelled pedestrian, its nose or the outer border of the shoulder. This is implemented by checking the coordinate in spaces of the pedestrians and evaluating if they are within the given radius of interaction. This can conservatively consider that if one agent has at least one part of its body within the social distance radius, it is assumed to be counted. The example in in Figure 1.2 shows one agent $i$ and one agent $j$ that are within a given social distance radius $R_i$. This assumption requires from a crowd model the information concerning the pedestrian trajectories over time (e.g. the parametric equations of pedestrian trajectories).

### 2.3. Exposure by face-to-face contact within a social distance radius

Occupant exposure in a confined space is assumed when people are in face-to-face contact to each other within a given angle of interaction in a social distance radius defined by the user. To facilitate implementation, the polar coordinates $\alpha_{ij}(t)$ and $\rho_{ij}(t)$ can be used for defining the position of the agent $j$ in the polar space in relation to each agent $i$ (see Figure 1.3), where $\rho_{ij}$ changes over time. Zero is the case in which people physically touch each other, the max value for $\rho_{ij} = R_i$ within the assumed social distance radius, so $\rho_{ij} = [0, R_i]$. $\alpha_{ij}(t)$ changes over time and it can vary from zero when the agent $j$ is in front of the agent $i$ to $\pm\pi$ when the agent $j$ is right behind the agent $i$. $\beta_{ij}(t)$ is the orientation of the agent $j$ in the polar space defined by the agent $i$. It can vary from zero when the agent $j$ is facing the agent $i$ to $\pm\pi$ when the agent $j$ is turning its back on the agent $i$. As such, $\beta_{ij}(t)$ can be used to evaluate how many agents are at face-to-face contact within a social distance radius at a given time. This can conservatively assume that if one agent has at least one part of its body within the social distance radius, it is assumed to be checked for the face-to-face contact criteria. This assumption requires from a crowd model the information concerning both the pedestrian trajectories over time (considering a given position within the simulated agent, e.g., the centre of the agent), as well as the direction of movement of each pedestrian (in order to know the face orientation). The user would then need to make assumptions concerning the angles leading to face-to-face contact between pedestrians.

### 2.4. Exposure by being in the same room

Occupant exposure in a confined space is here assumed when agents are in the same room/compartment (see Figure 1.4). This assumption requires the information from a crowd model concerning the number of building occupants in each room at each time-step.

### 2.5. Exposure by being in the same building

Occupant exposure in a confined space is here assumed when agents are in the same building (see Figure 1.5). This is the simplest assumption from a crowd modelling implementation standpoint, as it only requires the information concerning how many people are in the building in a given time-step.

## 3. EXPOSED: the occupant exposure model

In this work, we present a novel occupant exposure model, named EXPOSED. This model aims at estimating a set of metrics concerning occupant exposure in confined spaces. The assumptions underpinning the definition of these metrics is provided in the following paragraphs. It is worth mentioning that given the current lack of a comprehensive understanding of the SARS-CoV-2 virus transmission, we propose here a flexible and simple model to encourage its immediate usage and exploitation in the crowd modelling community.

Considering that there is no information available on the initial number of agents who are susceptible, infected, or recovered, the model aims at quantifying the exposure of the pedestrians in a confined space. The general formulation of EXPOSED is here presented considering different types of occupant exposure introduced in Section 2. In other words, the equations provided in this section can be applied for the five types of exposure defined in Sections 2.1-2.5.

Assuming that each agent $i$ can be exposed to a certain number of occupants based on the exposure assumption in use (see Section 2), it is possible to obtain the information concerning number of agents to which the agent $i$ is exposed to at each time-step $\boldsymbol{t} = [\, t_0, t_1, \ldots, t_q, \ldots, t_f \,]$ until it has left the building at the final time $t_f$. In this formulation we assume that the time-steps have the same magnitude (i.e., $t_{q+1} - t_q = \Delta t = constant \; \forall q$). This information can be represented as a set $\boldsymbol{E^i}$ representing the number of people each agent is exposed to at each time-step $t_q$.

$$\boldsymbol{E^i} = \{e_{t0}, e_{t1}, \ldots, e_{tq}, \ldots, e_{tf}\} \quad \forall\, i$$

[Equation 1]

where $e_{tq}$ is the $k$ number of agents $j$ to which each individual agent $i$ is exposed at the time-step $t_q$.

The information concerning each occupant exposure at each time-step can be represented in the form of the matrix $\boldsymbol{E_t^i}$ in Equation 2.

$$\boldsymbol{E_t^i} = \begin{pmatrix} E^1 \\ \vdots \\ E^i \\ \vdots \\ E^n \end{pmatrix} = \begin{pmatrix} e_{t0}^1 & \cdots & e_{tq}^1 & \cdots & e_{tf}^1 \\ \vdots & \ddots & \vdots & \ddots & \vdots \\ e_{t0}^i & \cdots & e_{tq}^i & \cdots & e_{tf}^i \\ \vdots & \ddots & \vdots & \ddots & \vdots \\ e_{t0}^n & \cdots & e_{tq}^n & \cdots & e_{tf}^n \end{pmatrix}$$

[Equation 2]

Since $\boldsymbol{E_t^i}$ presents the number of people each agent is exposed to at each $t_q$, it is therefore possible to obtain the information on the time $T_k^i$ each $i$ agent has been exposed to a given number of agents $k$ (i.e. the exposure time to 0 occupants, 1 occupant, ..., m occupants) by summing $t_q^i$, i.e., the number of time-steps $t_q$ in which each agent $i$ was interacting with a discrete number of people $k$, see Equation 3. The model user could assume that the exposure is considered either for any $t$ or only counting the time-steps in case of a minimum exposure time (e.g. counting the $t_q^i$ if the exposure last at least for a given number of seconds/minutes, i.e. a certain number of consecutive $t_q^i$ are required). The maximum number of occupants that agents are exposed to correspond to a maximum of $n - 1$ if the number of people in the building is restricted (i.e. if a maximum number of people is allowed in the building at the same time) or to a generic number of people $m$ if we assume a transient space.

$$T_k^i = \sum_{t_0}^{t_f} t_q^i \ \forall \ i, k$$

[Equation 3]

Considering the total time $t_f$ spent by all $n$ agents in the building, it is therefore possible to obtain a set of distributions $T_k$ of exposure times corresponding to a given discrete number of agents $k \geq 0$. Using a distribution from the two-parameter family of continuous probability distributions, $T_k$ can be defined by its mean ($\mu_k$) and standard deviation ($\sigma_k$) as shown in Equation 4.

$$T_k \ (\mu_k, \sigma_k^2)$$

[Equation 4]

These distributions provide useful information concerning the occupant exposure, including:
1) Maximum number of agents that people are exposed at the same time $m$
2) Longer time of exposure to each given number of agents $t_{k,max}^i$
3) Average $\mu_k$ and variance $\sigma_k^2$ of time spent in contact with a given number of agents and how those times are spread among contacts with different number of agents.

The values reported in these distributions - corresponding to each $k$ - range from a minimum possible value corresponding to no exposure (i.e. zero exposure) to a maximum time of exposure $t_{k,max}^i$ for each simulated agent $i$ to each number of $k$ agents they are exposed to. The summation over the data-points available for each of the values obtained $\forall \ k$ (see Equation 5) helps performing an assessment of the cumulative exposure $C_k$ to a given number of people $k$. The higher is the value of the summation, the greater is the occupant exposure for $k > 0$. The value of the summation for $k = 0$ is an indicator of how long people have not been exposed to other agents in the building; this is called here $C_0$.

$$C_k = \sum_{i=1}^{n} T_k^i$$

[Equation 5]

The sum of all $C_k$ with $k > 0$ provides a global assessment of exposure $G$ for the total time $t_f$ spent by all $n$ agents in the building (see Equation 6). To obtain $G$, each $C_k$ is multiplied by a factor $\gamma_k$ which increases the exposure in relation to the value of $k$. For instance, $\gamma_k$ can be assumed equal to 1 for $k = 1$ and with increasingly higher values when $k > 1$. The choice of the values for $\gamma_k$ is left to the model user.

$$G = \sum_{k=1}^{m} \gamma_k \, C_k$$

[Equation 6]

The model user can therefore obtain different values for $G$ in relation to the assumptions adopted for occupant exposure (see Section 2).

It should be noted that the matrix in Equation 2 can also be weighted considering the vulnerability of each individual occupant (e.g. considering their individual properties, such as age, physical abilities, etc. which can often be represented within a microscopic crowd model).

## 4. Coupled use of EXPOSED and a crowd model

The methodology for the use of EXPOSED together with a crowd model is presented in Figure 2 as a set of steps to follow.

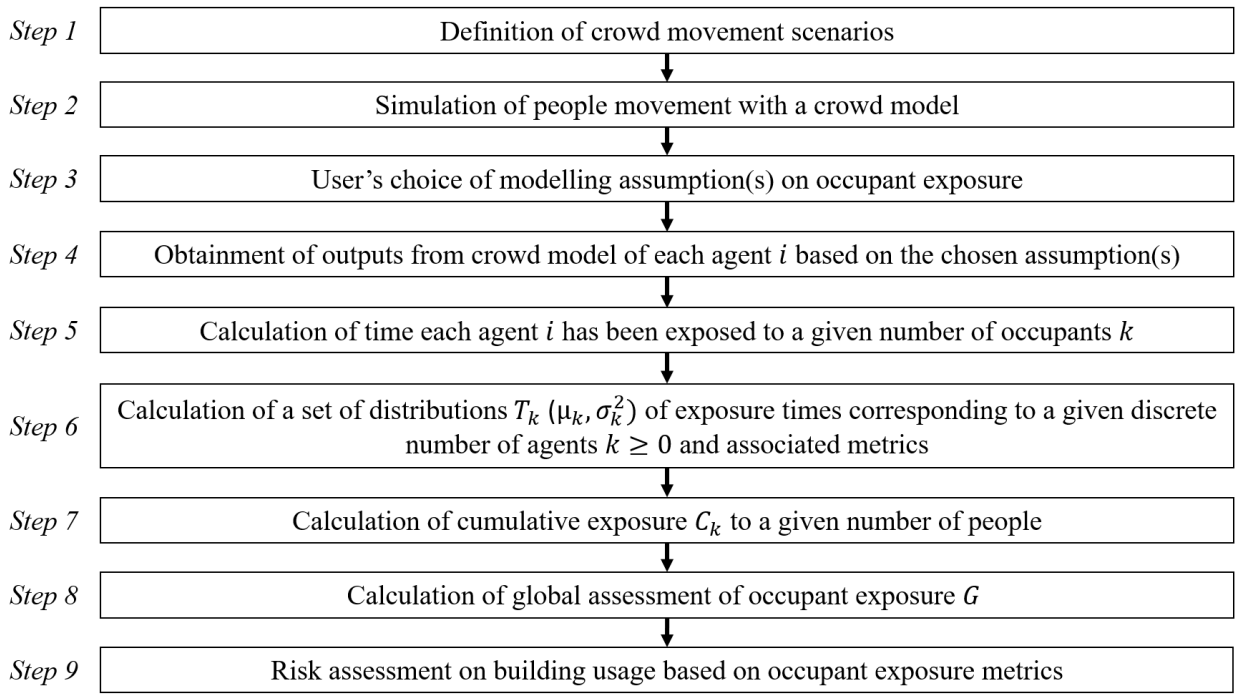

*Figure 2. Steps to be followed for the coupled use of the occupant exposure model EXPOSED and a crowd model.*

Step 1 is the definition of crowd movement scenarios. The user would have to set the assumptions on crowd movement and behaviour, such as circulation paths, group behaviour (intended here as a group of people moving together (Adrian et al., 2019), building occupant interactions, etc. Crowd movement and behaviour can be represented with different levels of sophistication in relation to the crowd model in use (Duives et al., 2013). The second step is the simulation of people movement using a crowd model selected by the user. After the simulations have been performed, the EXPOSED model user needs to decide which modelling assumption(s) to use concerning occupant exposure (Step 3) - as discussed in Section 2 of this paper - and obtain the needed crowd model outputs accordingly (Step 4). Depending on the type of disease transmission, the outputs required can include the trajectories of the agents, their direction of movement, agent dimensions, or the location of agents in the room/building. Step 5 includes the calculation of the occupant exposure time, i.e., the time $T_k^i$ each agent $i$ has been exposed to a given number of occupants $k$. It is therefore possible to calculate a set of distributions $T_k$ ($\mu_k, \sigma_k^2$) of exposure times and their associated metrics (Step 6). It is therefore possible to calculate the cumulative exposure $C_k$ to a given number of people (Step 7) and the global assessment of occupant exposure $G$ (Step 8). The occupant exposure metrics can then be used to perform risk assessments on building usage and subsequently inform decision making (Step 9).

## 5. Case Study

In this section a fictitious case study is presented to provide an example of the applicability of the EXPOSED model. The scenarios and hypotheses included in this case study are deliberately kept simple given its explanatory purpose. The values in use have been created with the Mersenne Twister pseudo-random number generator (Matsumoto & Nishimura, 1998).

A generic building layout with a fixed number of people able to access it equal to 10 is considered. Assuming that we obtained the required information concerning the simulated pedestrian movement from a crowd model, the exposure time is calculated with any of the five assumed criteria listed in Section 2. In this fictitious example, an exposure time between 0 and 5 min for each individual agent $i$ to a given number of agents $k$ in the building has been obtained with a pseudo-random number generator (presented in Appendix A), i.e. we are starting the application of the methodology from Step 5 of Figure 2.

Equation 3 can be represented for each agent $i$ as in Figure 3 (representing the exposure time of each agent $i$ to the other 9 people in the building, assuming that all cases of interactions occur).

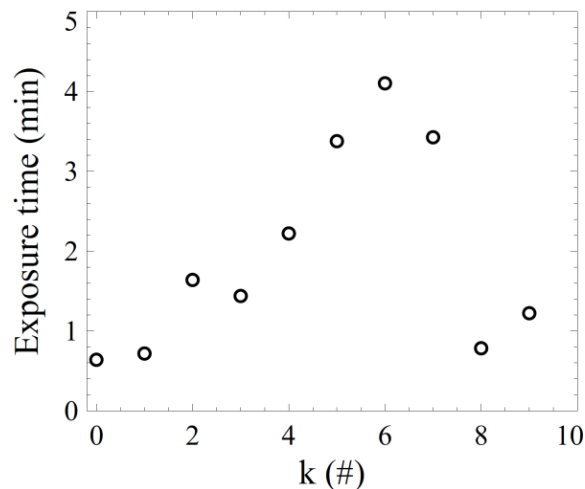

*Figure 3. Exposure time of an individual agent i to a given number of agents k in the building.*

All data-points ($\forall\ i = n$) as in Figure 3 can be obtained to show the exposure time of each agent to the other agents in the simulation $0 \leq k \leq n - 1$). Those can be visualized as boxplots (see Figure 4).

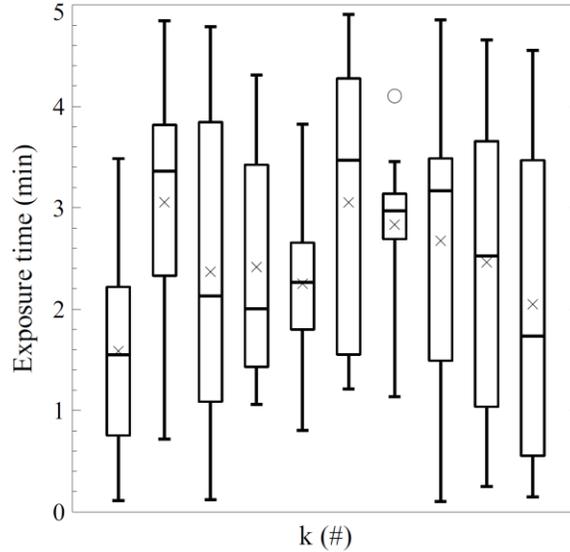

*Figure 4. Boxplots of exposure times of all $n$ agents to the other $n - 1$ agents in the building (k from 0 to 9 from left to right).*

Equation 4 is therefore associated with a given set of distributions $T_k$ having a certain average, maximum and standard deviation (see Table 1).

*Table 1. Average, standard deviation and maximum based on the distributions of exposure times $T_n^k$ for all occupants in the building and a given number $k$ of occupants they are exposed to (times are here reported in seconds).*

| k (#) | Average (s) | Standard deviation (s) | Max (s) |
|---|---|---|---|
| 0 | 95 | 62 | 209 |
| 1 | 183 | 80 | 291 |
| 2 | 142 | 95 | 287 |
| 3 | 145 | 70 | 259 |
| 4 | 135 | 46 | 229 |
| 5 | 183 | 82 | 294 |
| 6 | 170 | 45 | 246 |
| 7 | 160 | 98 | 291 |
| 8 | 148 | 90 | 279 |
| 9 | 123 | 93 | 273 |

This information is useful to identify the maximum number of people that agents are exposed at the same time (this is in this case the maximum number possible, i.e., $n - 1 = 9$, given the data generated in the example). In addition, the average and standard deviations of the time of exposures to each number of people is a useful information for performing a quantitative risk assessment.

It is now possible to calculate $C_k$ (including $C_0$) based on the summation of the data-points in the distributions (see Equation 5). The calculated $C_k$ values are presented in Table 2. It should be noted that in this case study, Table 2 presents the $C_k$ values considering $\gamma_k = 1$ or $a_k$ linearly increasing with $k$ (i.e. $\gamma_k = 2$ when $k = 2, \gamma_k = 3$ when $k = 3, \ldots, \gamma_k = n - 1 = 9$ when $k = 9$).

*Table 2. Calculated values of the integrals of the exposure curves (approximations made at 1 decimal).*

|  | $C_0$ | $C_1$ | $C_2$ | $C_3$ | $C_4$ | $C_5$ | $C_6$ | $C_7$ | $C_8$ | $C_9$ |
|---|---|---|---|---|---|---|---|---|---|---|
| k=1 | 15.9 | 30.5 | 23.7 | 24.2 | 22.5 | 30.5 | 28.3 | 26.7 | 24.6 | 20.5 |
| k increases linearly | 15.9 | 30.5 | 47.4 | 72.5 | 90.0 | 152.7 | 170.0 | 187.2 | 196.9 | 184.4 |

Finally, it is possible to obtain the global assessment of occupant exposure $G$ using Equation 6 (considering both assumptions of $\gamma_k = 1$ or $\gamma_k$ increasing linearly. The final value for $G$ when $\gamma_k = 1$ is 231.6 min, while assuming $\gamma_k$ increasing linearly $G$ corresponds to 1131.6 min. The calculated value for $G$ can be evaluated against an acceptance criteria or several $G$ values can be calculated to compare different scenarios, assumptions, or conditions.

## 6. Discussion

EXPOSED is a simple model which is designed to retrofit existing microscopic crowd models to perform the assessments of occupant exposure. This represents an invaluable information to perform risk assessments during pandemics in confined spaces such as buildings and transportation terminals. The outputs obtained with EXPOSED can be useful to perform several types of analysis and it can be applied for several types of assumptions concerning disease transmissions leading to a given exposure. EXPOSED is model-agnostic, as it is designed to exploit the existing capabilities of any microscopic crowd models able to simulate pedestrian movement and provide basic outputs concerning the location of agents in the simulated building over time. Those outputs are commonly calculated by most commercial and research crowd movement models (Duives et al., 2013; Gwynne et al., 1999).

One of the great advantages of EXPOSED is that it is not a new microscopic model of occupant exposure, but it can be used either as a post-processor of an existing crowd model or it can be implemented directly within an existing crowd model. To date, over 70 models are already available in the crowd modelling market, thus it was deemed appropriate to develop a model to retrofit existing tools (as they have already a large base of users (Lovreglio et al., 2019) and many have gone through dedicated verification and validation testing (Ronchi et al., 2014)) rather than developing a brand new model. Its flexibility along with its simplicity is deemed to encourage its immediate usage and exploitation in the crowd modelling community. This is particularly important during new pandemics in which new knowledge is constantly available and there may be a need to update the assumptions performed on disease transmission. In fact, the outputs provided by EXPOSED allow users to perform risk assessments of disease spreads in confined spaces as a function of the assumptions adopted for occupant exposure.

EXPOSED can be used both to evaluate crowd management solutions as well as for building design. During pandemics, crowd management solutions can be implemented within buildings to ensure that social distances are kept by building occupants and the risk of disease transmission is minimized. The results of EXPOSED can be used to quantitatively compare the effectiveness of different crowd management solutions and identify the ones which yield the lowest occupant exposure (i.e. the one corresponding to the lowest value of $G$ defined in Equation 6). EXPOSED gives also the opportunity to obtain a quantitative understanding on the impact of (e.g., crowd management) solutions aiming at influencing behavioural itineraries and crowd movement on occupant exposure in confined spaces.

Crowd models are currently designed for safety/comfort applications such as the optimisation of pedestrian flows or the calculation of evacuation times in case of a threat (Bellomo et al., 2016; Lovreglio et al., 2019; Ronchi, 2015). Therefore, their ability to investigate occupant exposure is limited, as they are not natively designed for this purpose. Previous attempts to implement capabilities for disease transmission applications with crowd modelling are either linked to specific assumptions on occupant exposure (Goscé et al., 2014) or they are based on assumptions which are generally not known to the modeller (i.e. how many people are initially infected in a building (D'Orazio et al., 2020). EXPOSED is deliberately designed to be a universal model which can be applied to any pedestrian crowd models regardless of their set of assumptions. It is also expected that new features of crowd models will be available as these tools can be used to simulate people movement during pandemics. This includes the representation of movement patterns, and social interactions during pandemics. It is expected that new crowd model features concerning change in movement and behaviours during pandemics would make the results obtained with EXPOSED more reliable and usable for risk assessment.

A possible application of EXPOSED is the possibility to compare the impact on occupant exposure while reducing/increasing the number of people in a building. This is a crucial type of information for policy makers, as during the re-opening phase of a pandemic, they often face the challenge to decide which type of people movement restrictions are the most appropriate for a given type of building. Similarly, EXPOSED can be used to evaluate how the occupant exposure changes in relation to the assumptions adopted for disease transmission. This is particularly important for the case of new pandemics (such as COVID-19) for which a clear understanding on the mechanism of disease transmission is not available yet (Lewis, 2020).

In the long term, EXPOSED could also be used to inform building design and minimize occupant exposure based on the geometric layout of a building. This can be particularly relevant for confined spaces in which a higher risk of disease transmission is expected (e.g. in the healthcare domain).

Future applications of EXPOSED could make use of the advances in crowd modelling (e.g. more detailed microscopic representations of people movement and behaviour during pandemics), thus including further mechanisms of transmission and more sophisticated interactions with the environment, e.g., physical contacts to objects rather than only considering people and areas. Current crowd models do several simplifications in the representation of the human body (calculations are generally made considering a circle/ellipse representing a human in a 2D space (Fruin, 1987) which is then extruded in a 3D environment). Novel crowd modelling approaches are investigating more accurate representation of human movement based on biomechanical models (Thompson et al., 2015), thus it is expected that they will be able in the future to investigate even more microscopic interactions with objects. This information could directly be exploited in risk assessments using EXPOSED.

An additional advantage on the use of microscopic modelling of crowd movement to assess occupant exposure is that it allows for in-depth considerations on the individual attributes of the agents. For instance, existing crowd models often implement socio-demographic variables (e.g. age, gender, physical abilities, functional limitations) (Duives et al., 2013; Geoerg et al., 2018) which affect people movement and in turn can affect occupant exposure. This means that the use of EXPOSED coupled with a microscopic crowd model takes into account the higher exposure of individuals whose attribute make them more vulnerable in a crowd (e.g. people walking slower or spending more time in performing way-finding in a building).

EXPOSED can also be used for the assessment of occupant exposure in buildings of the same type or different types. Policy makers can ensure that measures aiming at minimizing disease transmission result in a consistent level of occupant exposure. This type of modelling effort is deemed to facilitate the identifications of restrictions which provide a consistent level of disease transmission risk to building occupants. Similarly, EXPOSED can be used to perform a risk assessment of individual buildings. It is sensitive to assume that acceptance criteria corresponding to tolerable levels of exposure are identified (thresholds of tenability for buildings, in a similar fashion to what is currently performed in risk assessments for other building hazards (Meacham & Custer, 1995)) and used to evaluate the appropriate level of building usage.

The main limitations of the methodology proposed in this paper rely on the assumptions necessary for the calibration of both the underlying crowd model and EXPOSED. The verification and validation (V&V) of crowd models have been scrutinized through several research efforts over the years, including dedicated V&V procedures designed for different types of applications (IMO, 2016; Rimea, 2016; Ronchi et al., 2013). The outputs obtained with EXPOSED will greatly be affected by the validity of the underlying model used to simulate pedestrian movement. In addition, the EXPOSED model user would need to perform a set of assumptions to calibrate the model. The first key user input regards the assumptions for occupant exposure. Another important input to be identified is the value of the $\gamma_k$ multiplier (see Equation 6), as this can significantly affect the results. As more knowledge concerning the mechanisms of disease transmission and crowd behaviour during pandemics is available, the user would be able to appropriately calibrate those inputs.

## 7. Conclusion

This paper presents EXPOSED, a model designed to support risk assessment based on occupant exposure in confined spaces during pandemics. EXPOSED is model-agnostic as it is designed to make use of common outputs of existing microscopic crowd models. EXPOSED is designed to allow more informed decisions concerning building access restrictions during pandemics by performing a quantitative assessment of occupant exposure.

## Acknowledgements

Enrico Ronchi wishes to thank Cantene srl for providing financial support for the work presented in this paper. The authors thank Håkan Frantzich for providing valuable feedback on the work performed.

# Appendix A. Pseudo-random generated exposure times (in min) of all n agents to the other n-1 agents in the building

| $i$ | 1 | 2 | 3 | 4 | 5 | 6 | 7 | 8 | 9 | 10 |
|---|---|---|---|---|---|---|---|---|---|---|
| $k$ | | | | | | | | | | |
| 0 | 0.640 | 0.110 | 1.477 | 0.328 | 2.734 | 3.485 | 1.098 | 2.255 | 2.114 | 1.621 |
| 1 | 0.718 | 3.732 | 3.521 | 0.954 | 4.843 | 3.206 | 2.085 | 3.846 | 4.579 | 3.063 |
| 2 | 1.640 | 0.121 | 2.620 | 4.289 | 4.788 | 1.081 | 1.103 | 4.002 | 0.673 | 3.369 |
| 3 | 1.439 | 1.060 | 3.435 | 1.261 | 3.821 | 2.569 | 4.308 | 3.393 | 1.439 | 1.429 |
| 4 | 2.222 | 1.798 | 2.755 | 2.356 | 3.824 | 1.803 | 1.710 | 2.308 | 2.931 | 0.804 |
| 5 | 3.378 | 4.595 | 3.632 | 1.211 | 4.908 | 3.561 | 1.453 | 4.490 | 1.454 | 1.847 |
| 6 | 4.103 | 2.996 | 2.100 | 3.457 | 3.187 | 2.948 | 2.737 | 1.135 | 2.676 | 2.995 |
| 7 | 3.427 | 1.217 | 3.227 | 0.102 | 4.837 | 3.507 | 3.110 | 2.316 | 0.148 | 4.856 |
| 8 | 0.784 | 3.586 | 1.800 | 4.655 | 2.220 | 0.582 | 3.680 | 2.826 | 0.249 | 4.233 |
| 9 | 1.223 | 3.951 | 0.145 | 4.554 | 0.455 | 3.013 | 2.249 | 0.468 | 0.812 | 3.620 |